# Nonlinear study of local ballooning mode near the separatrix

T.F. Tang[1,2,*], X. Q. Xu[3], K. Li[4], M. Q. Wu[1,2], X. X. Zhang[4], X. Gao[1,2,4], G. Q. Li[4], T. Y. Xia[4], D. Z. Wang[5]

[1]*College of Physics and Optoelectronic Engineering, Shenzhen University, Shenzhen 518060, China*

[2]*Advanced Energy Research Center, Shenzhen University, Shenzhen 518060, China*

[3]*Lawrence Livermore National Laboratory, Livermore, CA 94550, United States of America*

[4]*Institute of Plasma Physics, Chinese Academy of Sciences, Hefei 230031, China*

[5]*School of Physics, Dalian University of Technology, Dalian 116024, China*

*\*Email: tang.tengfei.dut@gmail.com*

**Abstract**

Small edge-localized-mode (ELM), similar to the quasi-continuous exhaust (QCE), has been achieved by increasing the density at the separatrix. Starting from the Type-I ELM experimental data in EAST, we have performed a numerical separatrix density scan to study the formation of the small ELM using BOUT++ 6-field 2-fluid module. In the high separatrix density case, $n_{sep}/n_{ped} = 0.6$, localized collapse near the separatrix has been found. The corresponding ELM size, dominant mode number, and filament transport match the experimental observations of the QCE. Local ballooning mode near separatrix has been identified in the nonlinear simulation. The mode is driven by the local pressure gradient and the mode structure is constrained by the $E \times B$ shear.

## 1 Introduction

Large heat flux onto the divertor target is a critical issue for the future device, such as ITER and CFETR. Small heat flux width from the experimental scaling and theoretical prediction will further aggravate this issue. High separatrix density benefits the radiation power loss and divertor detachment, which could alleviate the heat flux problem. With higher separatrix density, the quasi-continuous exhaust (QCE) regime has been found in the ASDEX Upgrade (AUG) and TCV. The QCE could significantly reduce the heat flux and widens the heat flux width in the divetor target[1–3], which could be a possible solution to the heat flux issue in future devices.

QCE was called small edge-localized-mode (ELM) or Type-II ELM in previous studies (see the review paper of the small ELM [4]). The name QCE comes from the observation of enhanced filamentary transport at the plasma edge compared to inter Type-I ELM phases [3,5]. A hypothesis theory of the local ballooning mode near separatrix that triggers the QCE has been introduced via the local parameter analysis [6]. The resistive ballooning mode also could be the mode as the collisionality is high when separatrix density is high.

In this paper, we try to perform a numerical separatrix density scan to identify the local ballooning mode that trigger the QCE. The onset of the QCE requires the following plasma conditions: high triangularity, magnetic configuration close to double null, and high separatrix density. We start from Type-I ELM experimental data with high triangularity and a magnetic configuration close to double null in EAST. Then we artificially increase the separatrix density with fixed pedestal height and width and perform the nonlinear simulations to identify the local ballooning mode. The 6-field 2-fluid module in BOUT++ framework is used in this study, as BOUT++ framework is widely used in edge plasma simulations, especially in ELMs and pedestal turbulence studies [7–12].

The paper is organized as following: sec. 2 gives the parameters of the Type-I ELM experimental discharge, the plasma profiles, pedestal structure and the separatrix density setting is also given; the local analysis of the plasma and the linear BOUT++ simulation is shown in sec. 3; sec. 4 provides the detail of the nonlinear simulation and the fluctuation characteristics have been analyzed; the conclusion is given in sec.5.

## 2 Experimental discharge

### 2.1 Plasma profiles

One Type-I ELM shot in EAST, #55568 @ 3050ms, has been chosen to study the separatrix density effects. The discharge parameters are shown in Table I. The triangularity of this shot is 0.47, similar to the triangularity in AUG QCE experiments. And $\varDelta_{sep}$ is 5.6mm, close to the double null configuration, which is shown in figure 1 (a). The red dash curve in figure 1 (a) is the last closed flux surface, a lower single null configuration. The density and the temperature profiles are given in figure 1 (b) and (c). The ratio between the separatrix density and the pedestal top density is $n_{sep}/n_{ped} = 0.32$. The density profile in the pedestal is fitted by the following fitting function[13]:

$$n_{pedstal}(r) = \frac{a_{height} - a_{sol}}{2}\left[mtanh\left(\frac{2(a_{pos} - r)}{a_{width}}, a_{slope}\right) + 1\right] + a_{sol} \quad (1)$$

$$mtanh(x, a_{slope}) = \frac{(1 + a_{slope}x)e^x - e^{-x}}{e^x + e^{-x}} \quad (2)$$

where $a_{height}$ is the pedestal height, $a_{width}$ is the pedestal width, $a_{pos}$ is the location of the center of the edge transport barrier, $a_{sol}$ is the separatrix density and $a_{slope}$ represent a linearly rising profile in the core.

| Shot | Time (ms) | $I_p$ (kA) | $\beta_{p,ped}$ | $\langle n_e \rangle$ ($10^{19}$m$^{-3}$) | $\beta_N$ | $\beta_p$ | $\delta$ | $\Delta_{sep}$ (mm) | $B_T$ (T) | $P_{total}$ (MW) |
|---|---|---|---|---|---|---|---|---|---|---|
| 55568 | 3050 | 360 | 0.35 | 3.24 | 0.80 | 0.95 | 0.47 | 5.6 | 2.17 | 2.3 |

Table I discharge parameters of #55568 @ 3050 ms

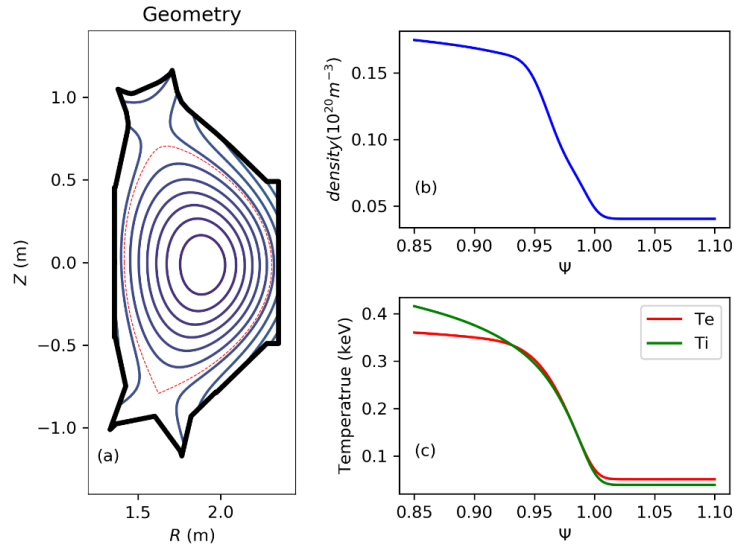

Figure 1 (a) magnetic geometry, (b) density profile, (c) temperature profile

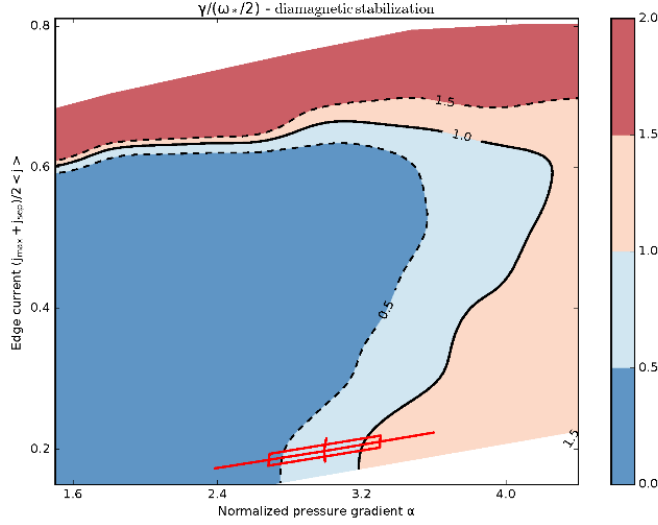

Figure 2 the peeling-ballooning diagram

The equilibrium is reconstructed by the kinetic EFIT code[14]. With this equilibrium, we first verify that this shot is at the ballooning branch of the peeling-ballooning stability diagram. To get the peeling-ballooning stability boundary of this shot, the VARYPED and ELITE code are used[15,16]. VARYPED code can generate a series of equilibrium by varying the pedestal pressure and current profiles[15]. And ELITE is an eigenvalue solver MHD code to calculate the growth rate of the peeling-ballooning mode[16]. The peeling-ballooning boundary is the black solid curve in figure 2. The experimental data point is the red square in figure 2, the error bar is set as 10%. According to the peeling-ballooning diagram, this shot is at the ballooning branch, and the ballooning mode is unstable for this shot.

**2.2 Pedestal structure and separatrix density scan**

The pedestal height and width have the separatrix density dependence [17]. So, we first studied the separatrix density effects on the pedestal height and width in this section. The REPED model is used here to calculate the pedestal height and width. REPED model is similar to the EPED model but with the real experimental geometry[18,19]. In the EAST pedestal structure study, the default $n_{sep}/n_{ped}$ is set as 0.25, and show good agreements with experimental measurements[18]. By artificially increase the $n_{sep}/n_{ped}$, the pedestal heights and widths of this shot are calculated via REPED model and shown in figure 3. No significant degradation on the pedestal height and width has been found in this separatrix density scan, and the difference is smaller than 7%. Therefore, we could assume the pedestal height and width are fixed during the density scan. With fixed pedestal height and width, we change the separatrix density, $a_{sol}$ in equation (1), and get a series density profiles, shown in figure 4.

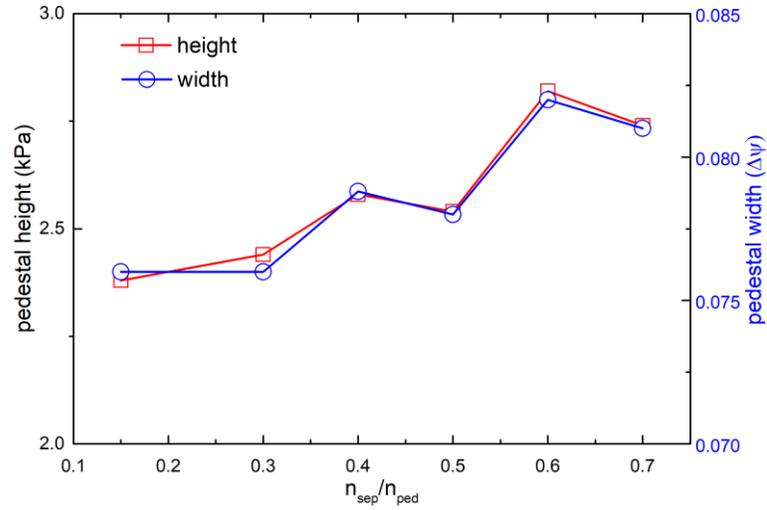

Figure 3 pedestal height and width under different separatrix density via REPED

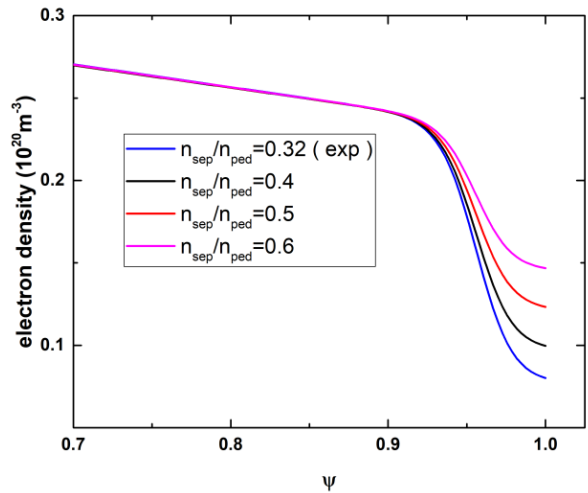

Figure 4 density profile with fixed height and width under different separatrix density

## 3 Local analysis and linear simulation

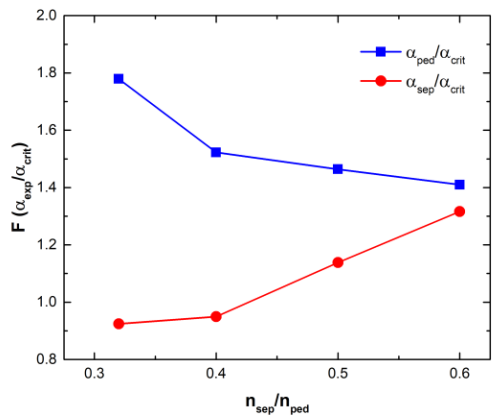

Figure 5 local normalized pressure gradient and the normalized separatrix pressure gradient ($\alpha_{crit}$ is calculated by Baloo code).

In this separatrix density scan, the temperature profile is fixed, corresponding normalized pressure gradient $\alpha$ inside the pedestal and at the separatrix will change accordingly. The normalized pressure gradient threshold of an unstable ballooning mode is calculated by the Baloo code[20], marked as $\alpha_{crit}$. The ratios between the normalized pressure gradient $\alpha_{exp}$ and the $\alpha_{crit}$ inside the pedestal and at the separatrix, $F(\alpha_{exp}/\alpha_{crit})$, are calculated and shown in figure 5. $\alpha_{ped}$ is the normalized pressure gradient at the peak gradient location. $F(\alpha_{ped}/\alpha_{crit})$ become smaller as the separatrix density increase, while $F(\alpha_{sep}/\alpha_{crit})$ at the separatrix increase, and is greater than 1 when $n_{sep}/n_{ped} > 0.5$. In the local $s - \alpha$ model, when $F(\alpha_{exp}/\alpha_{crit}) > 1.0$, the ballooning mode is unstable. From the local $F(\alpha_{exp}/\alpha_{crit})$ calculation by the Baloo code, there will be a local ballooning mode near the separatrix.

To verify the local ballooning mode near the separatrix, the BOUT++ linear simulation with 6-field 2-fluid module is conducted. The linear simulations under BOUT++ provide the growth rate and the mode structure of the fastest growing mode. The MHD growth rates from the toroidal mode number scan are shown in figure 6. The growth rate decreases when $n_{sep}/n_{ped}$ increases, a result of pressure gradient reduction inside the pedestal. The root mean square (RMS) of the pressure perturbation at the outer mid-plane (OMP) gives the mode structure, including the peak location and the radial extension of the mode. In figure 7, the pressure perturbations of $n = 30$ show the mode shifts outwards with the increasing separatrix density. The peak pressure gradient locations are also shifted outward, but shift distances of the peak pressure gradient location are different from shift distances of the mode. Peak mode location is between the peak gradient location and the separatrix. Check the radial extension of the perturbation, mode structures show no local mode, but mode peaks closer to the separatrix when separatrix density is higher.

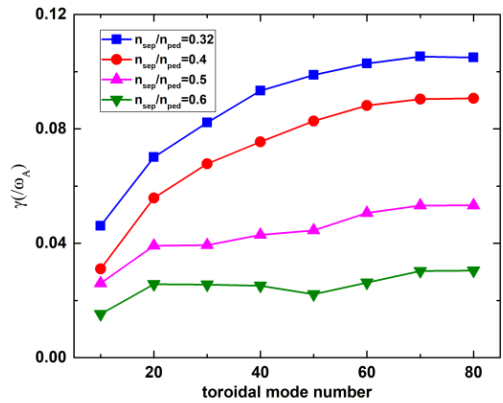

Figure 6 growth rates with different separatrix density

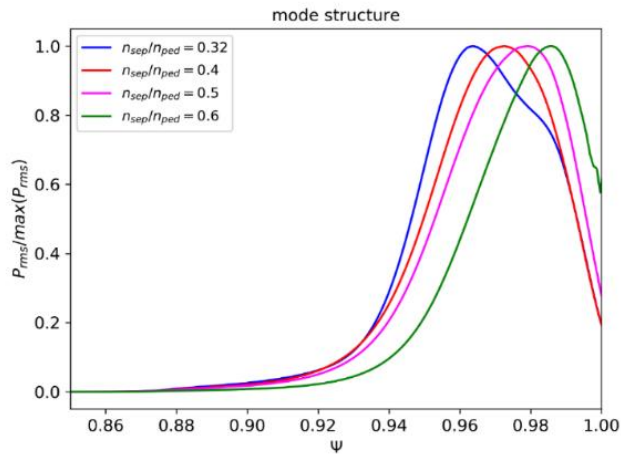

Figure 7 pressure perturbation of toroidal mode number n=30 with different separatrix density

The collisionality increases when separatrix density increases and resistive ballooning mode could be unstable and become dominant mode near separatrix. To verify this assumption, the growths rate from the resistive MHD module are calculated and compared to the MHD results. As aforementioned, the Spitzer resistivity is used in the simulations. The growth rates of toroidal mode number n=30 are given in figure 8. The growth rates calculated from the resistive MHD module are similar to the results from the MHD module. Thus, in these cases, the resistive ballooning mode is almost stable, the ballooning mode is still the dominant mode in the pedestal.

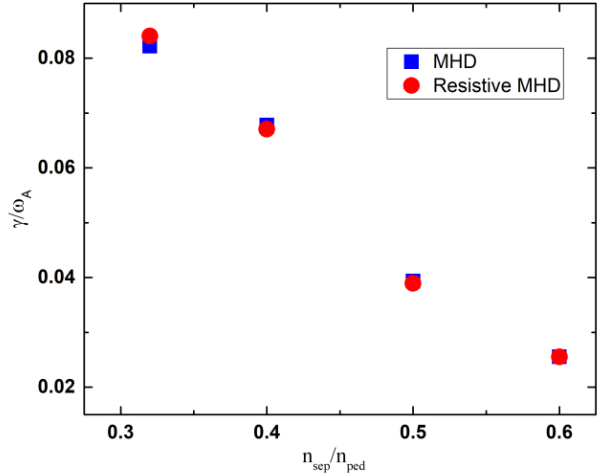

Figure 8 Resistive MHD growth rates versus MHD results, toroidal number n=30

## 4 Nonlinear simulations

### 4.1 Density perturbation

The purpose of this paper is to get the small ELM (QCE) via artificially increasing the separatrix density and to understand the formation of the QCE. To this end, the nonlinear separatrix scan has been done. In figure 9 (a), we show the RMS density fluctuations of the three nonlinear simulations. Due to the numerical difficulties, the simulation times of these three simulations are different, but all simulations have a steady saturation phase. With higher separatrix density, the period of the linear phase is longer and the saturation level is lower. The original density profile and the density profile at the time slice $t = 800\ \tau_A$ of three cases are given in figure 9 (b). Different from the $n_{sep}/n_{ped} = 0.32$ case and $n_{sep}/n_{ped} = 0.5$ case, $n_{sep}/n_{ped} = 0.6$ case has no visible crash inside the pedestal, only a small collapse of the profile near separatrix. The toroidal averaged perturbation or dc component of the perturbation at the OMP provides more details of the profile collapse. The comparison of dc component at the OMP between the $n_{sep}/n_{ped} = 0.32$ case and $n_{sep}/n_{ped} = 0.6$ case in figure 9 (c) and (d) shows the $n_{sep}/n_{ped} = 0.32$ case has a typical Type-I ELM crash characteristics, that the collapse of the profile is near the peak gradient location and moves inward due to the gradient reduction, while the $n_{sep}/n_{ped} = 0.6$ case has a small crash near the peak gradient location, and a more localized collapse near the separatrix can be found. This phenomenon is similar to the QCE observation and the local ballooning mode theory raised by the QCE explanation.

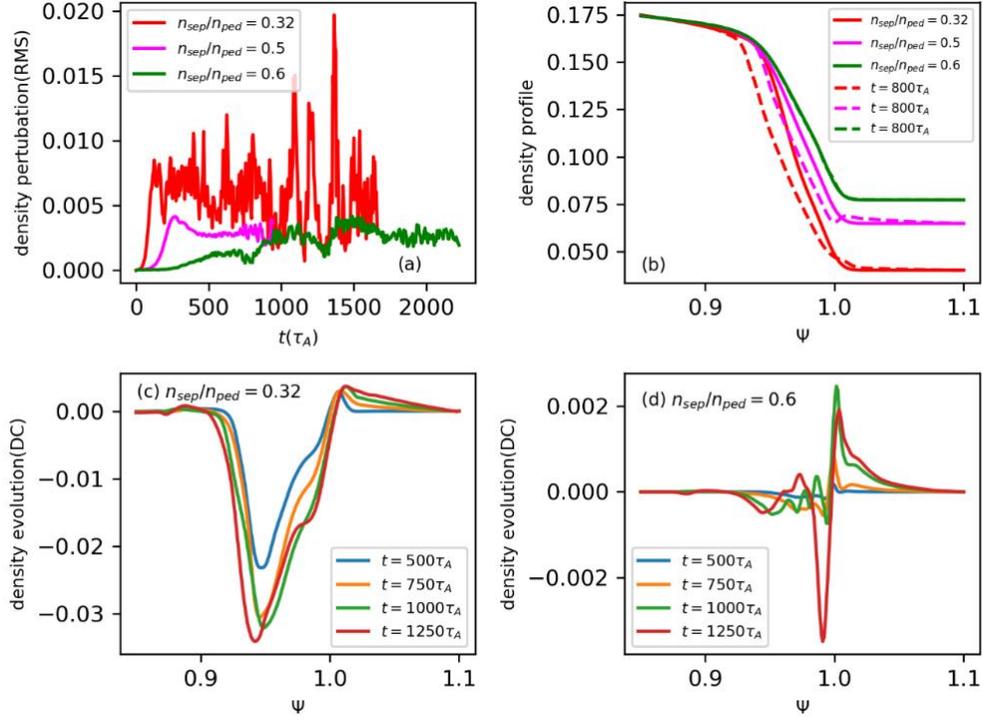

Figure 9 (a) density perturbation; (b) profile evolutions; (c) dc component of $n_{sep}/n_{ped} = 0.32$ case; (d) dc component of $n_{sep}/n_{ped} = 0.6$ case. (Normalized by $10^{20} m^{-3}$)

## 4.2 Small ELM characteristics

In this section, we use more data analysis to show the ELM characteristics of $n_{sep}/n_{ped} = 0.6$ case. First, The ELM size ($\Delta W/W_{ped}$, $\Delta E$) is calculated using the formula[21]:

$$\Delta E = \frac{\Delta W}{W_{ped}} = \frac{\int_{R_{in}}^{R_{out}} \oint dRd\theta (P_0 - <P>_\zeta)}{1.5 P_{ped} V_{plasma}} \quad (1)$$

Here $V_{plasma}$ is the plasma volume, and $P_{ped}$ is the pedestal pressure. The ELM sizes are given in figure 10. $n_{sep}/n_{ped} = 0.32$ case has a large ELM size of around 8%. When the separatrix density increases the ELM size decrease. As for the $n_{sep}/n_{ped} = 0.5$ case, the ELM size, larger than 2%, indicates the ELM is still large ELM. The $n_{sep}/n_{ped} = 0.6$ case has the smallest ELM size of 0.4%, a result of the localized crash near the separatrix, shown in figure 9 (d). This ELM size is smaller and comparable to the QCE ELM size or Type-II ELM size.

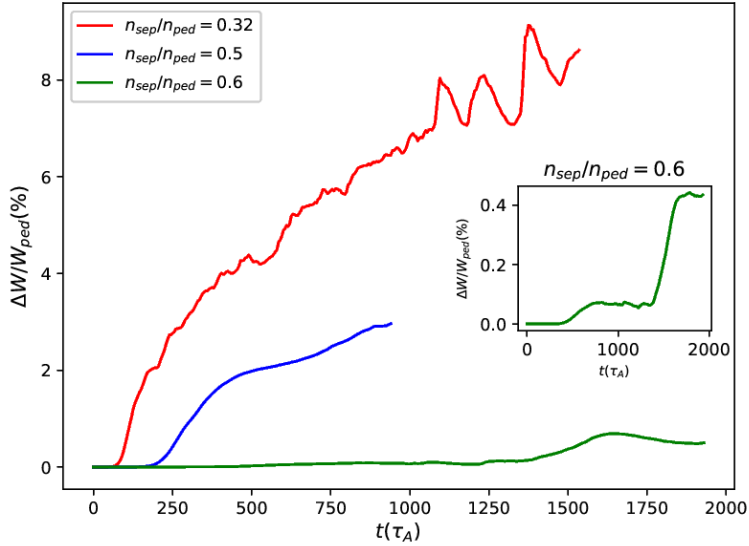

Figure 10 ELM size from the separatrix density scan

The profile collapse of the electron density and temperature is shown in figure 11. In the early stage during the crash process, the density and temperature profile evolutions are quite the same. In the later stage, the temperature profile crash becomes larger than the density profile crash. But no significant difference between them.

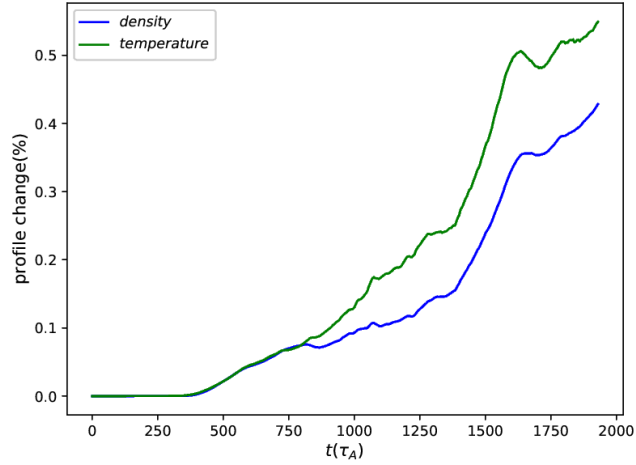

Figure 11 temperature and density profile change in the $n_{sep}/n_{ped} = 0.6$ case

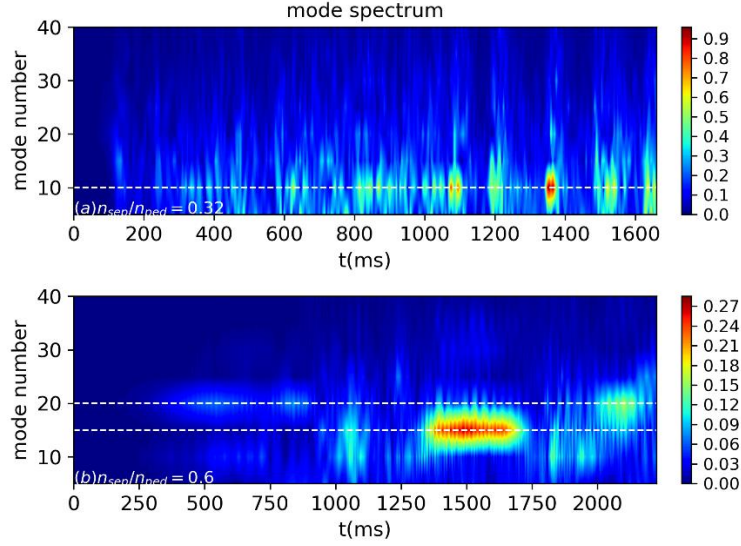

Figure 12 mode spectrum in the nonlinear simulation

The mode spectrums of $n_{sep}/n_{ped} = 0.6$ case and $n_{sep}/n_{ped} = 0.32$ case are calculated and given in figure 12. The signal is the perturbation of the parallel magnetic potential $A_\parallel$ near the separatrix. The dominant mode number of $n_{sep}/n_{ped} = 0.32$ case is $n = 10$ and $n_{sep}/n_{ped} = 0.6$ case has a two dominant mode number $n = 15$ and $n = 20$. In AUG and MAST Type-II studies, the toroidal mode number of these ELMs is a factor of 2 higher compared to type-I ELM [22], this shares the same feature.

Another interesting observation is the filaments from the pedestal to the SOL [5]. In the experimental study, the emission change of one neutral helium line was used to investigate the filaments structures [5]. The intensity of the emission line depends on both density and temperature. For simplicity, we only analysis the density fluctuation. The evolution of the density fluctuation at the OMP is shown in figure 13, the white dash line is the separatrix. Similar to the observation in reference [5], the density filaments, generated inside the separatrix, move into SOL, detached from the bulk plasma inside the separatrix in the later time.

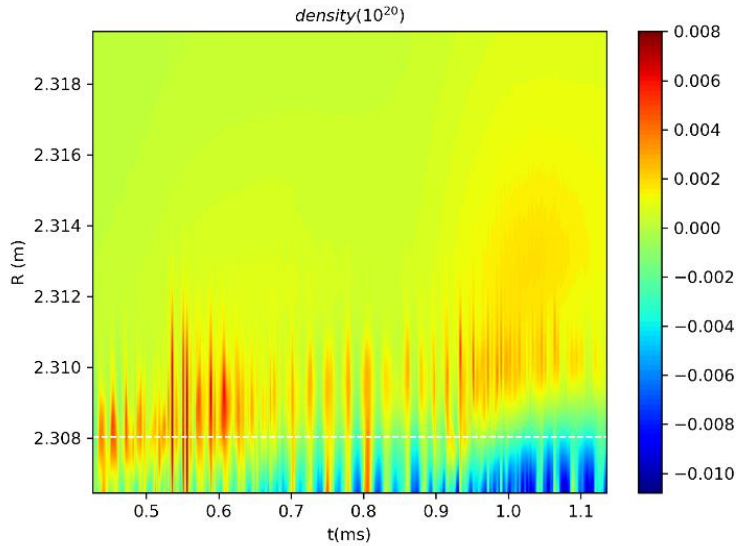

Figure 13 filaments across the separatrix in $n_{sep}/n_{ped} = 0.6$ case (the white dash line is the separatrix)

### 4.3 Mode structure and formation mechanism

In this section, the detail of the pressure perturbations is analyzed. We focus on the mode structure and mode spectrum to understand the formation of the small ELM. In figure 14, the pressure perturbations of the $n_{sep}/n_{ped} = 0.32$ case and $n_{sep}/n_{ped} = 0.6$ case at the OMP are given. The white dash line is the separatrix and the white solid curve is the peak pressure perturbation location. The $n_{sep}/n_{ped} = 0.32$ case show the fluctuations originally generated near the peak pressure gradient location move inward. The ELM crash causes the local flattening of the pressure gradient and the peak pressure gradient location moves inward, leading the fluctuations to move inward. As for the $n_{sep}/n_{ped} = 0.6$ case, the fluctuation peak oscillates between the peak pressure location and the separatrix.

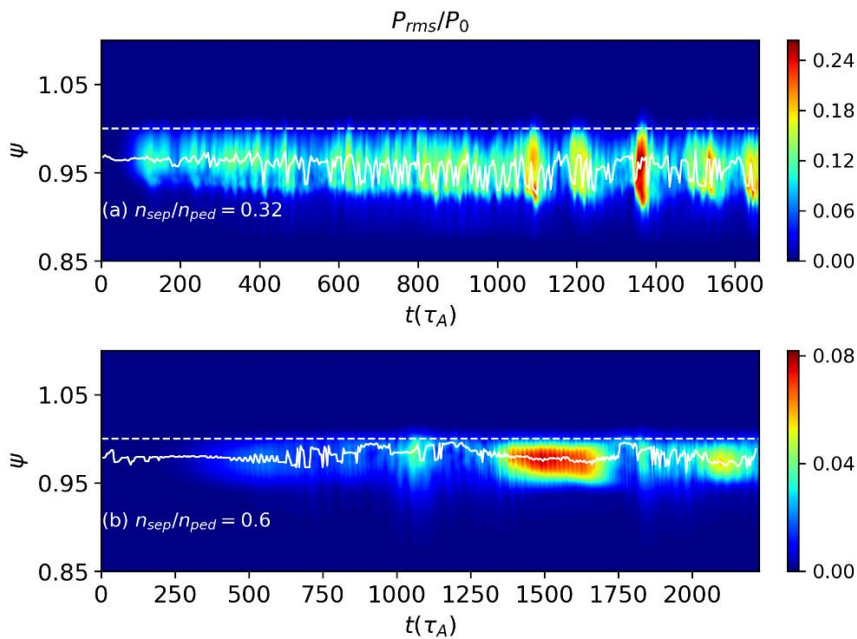

Figure 14 Pressure perturbation at the OMP of two cases (the white solid curve is the peak pressure perturbation location, and the white dash line is the separatrix)

In the linear simulation, only the fastest growing mode can grow up, while the sub-dominant mode will not. In the nonlinear simulation, the sub-dominant mode can grow up at the same time. In figure 15, the DC-part of the perturbation at the OMP is given. The red dash line indicates the separatrix. Two 'holes' along the radial direction can be found in figure 15 (a). Each 'hole' represents one mode there, then, there are two modes at this time slice. The inner mode is the one inside the pedestal, similar to the Type-I ELM, driven by the pressure gradient inside the pedestal. If there is only one global mode in the nonlinear simulation, the mode location will be where the 'hole' and 'blob' are generated at the same time, thus, the location is where the DC-part of the perturbation is zero. However, with two modes overlapped radically, we cannot tell where the mode locations are in this plot. In the later time of $t = 1200\tau_A$, the outer mode becomes the dominant mode.

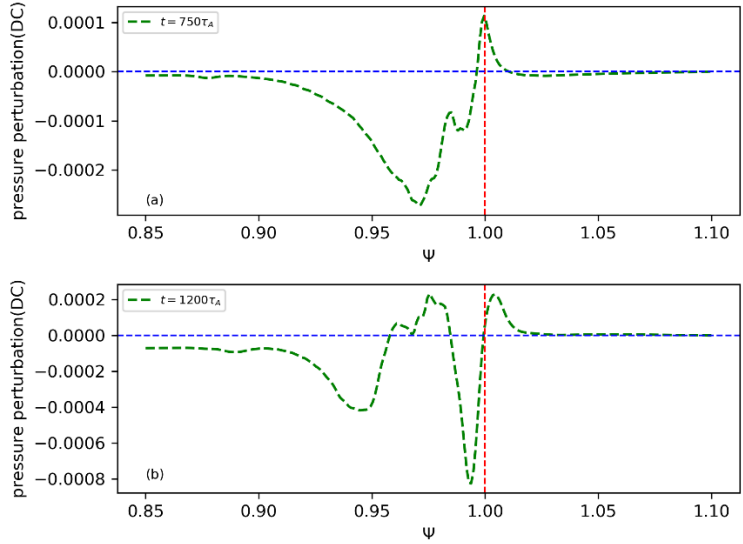

Figure 15 DC-part of the perturbation at the OMP

To further understand the outer mode, we do a fast Fourier transformation (FFT) and get the perturbation of each toroidal mode number n in our simulation. The mode number of the dominant mode (outer mode) is $n = 15$, and the perturbations of this mode at the OMP are given in figure 16 (green curve). The perturbation of the most unstable mode in linear simulation is also given for comparison, which is the red curve in figure 16. The $E \times B$ shearing rate profile, calculated by the formula

$$\omega_{E \times B} = \frac{(RB_\theta)^2}{B} \frac{\partial}{\partial \psi}\left(\frac{E_r}{RB_\theta}\right) \qquad (2)$$

is also given in figure 16. The peak perturbation location of the linear mode and the outer mode are consistent with the location where $E \times B$ shearing rate is zero. A result of the stabilizing effects of the $E \times B$ shear. Therefore, the nature of the outer mode can be explained as follow: with higher density near separatrix, the local pressure gradient exceeds the local ballooning mode threshold and triggers a ballooning mode near the separatrix. The radial structure of this mode is constrained by the $E \times B$ shear. The $E \times B$ shearing rate profile is affected by the density profile via the force balance equation of the electrical field. The outer mode is finally located between the separatrix and the peak pressure gradient location inside the pedestal where the $E \times B$ shearing rate is zero. The oscillation region in figure 14 (b) coincides with the $E \times B$ shearing rate well in figure 16.

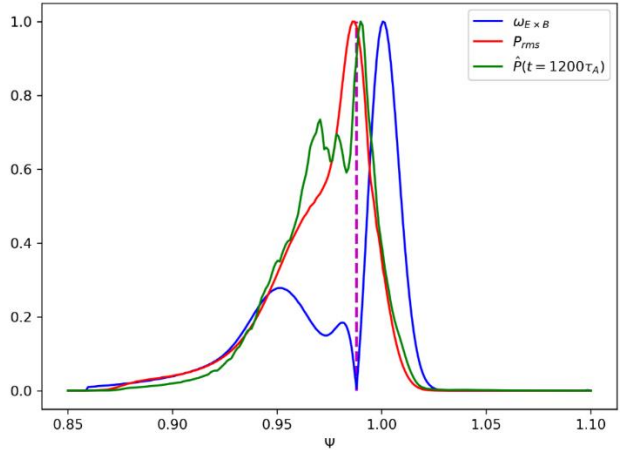

Figure 16 ExB shear and the mode structure

## 5 Conclusion

In this paper, we have conducted a separatrix density scan to identify the local ballooning mode that triggers the QCE. With the separatrix density increases, the local pressure gradient near the separatrix increases and exceed the local ballooning threshold, while the pressure gradient inside the pedestal decreasing, leading to a small linear growth rate. The most unstable mode in the linear simulation peaks a location between the separatrix and the peak pressure gradient location. The resistive ballooning mode has no impact on this shot.

In the nonlinear simulation, the high separatrix density case with $n_{sep}/n_{ped} = 0.6$ has a localized profile collapsed near the separatrix. The ELM size of this case is smaller than the low-density case, even smaller than 1%. The dominant mode comparison and filaments structure near the separatrix match the experimental observations of the QCE. Further fluctuation analyses identify the local ballooning mode near the separatrix. And the $E \times B$ shear constrains the mode structure and profile collapse in the simulations.

The grassy ELM also has similar characteristics as the QCE, such as high separatrix density and high triangularity[12,23]. The difference between QCE and grassy ELM is the 20-70 kHz fluctuation on the density and magnetic signal at $\rho \sim 0.75 - 0.9$. However, the profile in the core is from the TYPE-I ELM experiment, and no 20-70 kHz fluctuation can be found in the simulations. Therefore, we cannot tell whether it's QCE or grassy ELM in this paper. Nevertheless, the local ballooning mode could exist and trigger the small ELM near separatrix is verified in this paper. Future works will use the experimental data from QCE or grassy ELM to further verify this.

**Acknowledgment**

The authors would like to thank Dr. B. Zhu, N. M. Li and T. Zhang for useful discussions and the EAST team for providing the experimental data. This work was also supported by the National Key R&D Program of China Nos. 2017YFE0301206, 2017YFE0300402 and 2017YFE0301100. This work was also performed for USDOE by LLNL under DE-AC52-07NA27344 and LLNL-JRNL-823052.

**Reference**

[1] Dunne M G, Potzel S, Reimold F, Wischmeier M, Wolfrum E, Frassinetti L, Beurskens M, Bilkova P, Cavedon M, Fischer R, Kurzan B, Laggner F M, McDermott R M, Tardini G, Trier E, Viezzer E and Willensdorfer M 2017 The role of the density profile in the ASDEX-Upgrade pedestal structure *Plasma Phys. Control. Fusion* **59** 014017

[2] Labit B, Eich T, Harrer G F, Wolfrum E, Bernert M, Dunne M G, Frassinetti L, Hennequin P, Maurizio R, Merle A, Meyer H, Saarelma S, Sheikh U, Adamek J, Agostini M, Aguiam D, Akers R, Albanese R, Albert C, Alessi E, Ambrosino R, Andrèbe Y, Angioni C, Apruzzese G, Aradi M, Arnichand H, Auriemma F, Avdeeva G, Ayllon-Guerola J M, Bagnato F, Bandaru V K, Barnes M, Barrera-Orte L, Bettini P, Bilato R, Biletskyi O, Bilkova P, Bin W, Blanchard P, Blanken T, Bobkov V, Bock A, Boeyaert D, Bogar K, Bogar O, Bohm P, Bolzonella T, Bombarda F, Boncagni L, Bouquey F, Bowman C, Brezinsek S, Brida D, Brunetti D, Bucalossi J, Buchanan J, Buermans J, Bufferand H, Buller S, Buratti P, Burckhart A, Calabrò G, Calacci L, Camenen Y, Cannas B, Cano Megías P, Carnevale D, Carpanese F, Carr M, Carralero D, Carraro L, Casolari A, Cathey A, Causa F, Cavedon M, Cecconello M, Ceccuzzi S, Cerovsky J, Chapman S, Chmielewski P, Choi D, Cianfarani C, Ciraolo G, Coda S, Coelho R, Colas L, Colette D, Cordaro L, Cordella F, Costea S, Coster D, Cruz Zabala D J, Cseh G, Czarnecka A, Cziegler I, D'Arcangelo O, Dal Molin A, David P, et al 2019 Dependence on plasma shape and plasma fueling for small edge-localized mode regimes in TCV and ASDEX Upgrade *Nucl. Fusion* **59** 086020

[3] Faitsch M, Eich T, Harrer G F, Wolfrum E, Brida D, David P, Griener M and Stroth U 2021 Broadening of the power fall-off length in a high density, high confinement H-mode regime in ASDEX Upgrade *Nucl. Mater. Energy* **26** 100890

[4] Viezzer E 2018 Access and sustainment of naturally ELM-free and small-ELM regimes *Nucl. Fusion* **58** 115002

[5]     Griener M, Wolfrum E, Birkenmeier G, Faitsch M, Fischer R, Fuchert G, Gil L, Harrer G F, Manz P, Wendler D and Stroth U 2020 Continuous observation of filaments from the confined region to the far scrape-off layer *Nucl. Mater. Energy* **25** 100854

[6]     Harrer G F, Wolfrum E, Dunne M G, Manz P, Cavedon M, Lang P T, Kurzan B, Eich T, Labit B, Stober J, Meyer H, Bernert M, Laggner F M and Aumayr F 2018 Parameter dependences of small edge localized modes (ELMs) *Nucl. Fusion* **58** 112001

[7]     Dudson B D D, Umansky M V V., Xu X Q Q, Snyder P B B and Wilson H R R 2009 BOUT++: A framework for parallel plasma fluid simulations *Comput. Phys. Commun.* **180** 1467–80

[8]     Xu X Q, Dudson B, Snyder P B, Umansky M V. and Wilson H 2010 Nonlinear Simulations of Peeling-Ballooning Modes with Anomalous Electron Viscosity and their Role in Edge Localized Mode Crashes *Phys. Rev. Lett.* **105** 175005

[9]     Xia T Y and Xu X Q 2015 Nonlinear fluid simulation of particle and heat fluxes during burst of ELMs on DIII-D with BOUT++ code *Nucl. Fusion* **55** 113030

[10]    Tang T F, Shi H, Wang Z H, Zhong W L, Xia T Y, Xu X Q, Sun J Z and Wang D Z 2018 Quasi-coherent mode simulation during inter-ELM period in HL-2A *Phys. Plasmas* **25** 122510

[11]    Huang Y Q, Xia T Y, Xu X Q, Kong D F, Wang Y M, Ye Y, Qian Z H, Zang Q, Wu M P, Chu Y Q, Liu H Q, Gui B, Xiao X T and Zhang D Z 2020 Nonlinear simulation and energy analysis of the EAST coherent mode *Nucl. Fusion* **60** 026014

[12]    Xu G S, Yang Q Q, Yan N, Wang Y F, Xu X Q, Guo H Y, Maingi R, Wang L, Qian J P, Gong X Z, Chan V S, Zhang T, Zang Q, Li Y Y, Zhang L, Hu G H and Wan B N 2019 Promising High-Confinement Regime for Steady-State Fusion *Phys. Rev. Lett.* **122** 255001

[13]    Groebner R ., Baker D ., Burrell K ., Carlstrom T ., Ferron J ., Gohil P, Lao L ., Osborne T ., Thomas D ., West W ., Boedo J ., Moyer R ., McKee G ., Deranian R ., Doyle E ., Rettig C ., Rhodes T . and Rost J . 2001 Progress in quantifying the edge physics of the H mode regime in DIII-D *Nucl. Fusion* **41** 1789–802

[14]    Li G Q, Ren Q L, Qian J P, Lao L L, Ding S Y, Chen Y J, Liu Z X, Lu B and Zang Q 2013 Kinetic equilibrium reconstruction on EAST tokamak *Plasma Phys. Control. Fusion* **55** 125008

[15]    Osborne T H, Snyder P B, Burrell K H, Evans T E, Fenstermacher M E, Leonard A W, Moyer R A, Schaffer M J and West W P 2008 Edge stability of stationary ELM-suppressed regimes on DIII-D *J. Phys. Conf. Ser.* **123** 012014

[16]    Snyder P B, Wilson H R, Ferron J R, Lao L L, Leonard A W, Osborne T H, Turnbull A D, Mossessian D, Murakami M and Xu X Q 2002 Edge localized modes and the pedestal: A model

based on coupled peeling–ballooning modes *Phys. Plasmas* **9** 2037–43

[17]   Frassinetti L, Saarelma S, Verdoolaege G, Groth M, Hillesheim J C, Bilkova P, Bohm P, Dunne M, Fridström R, Giovannozzi E, Imbeaux F, Labit B, de la Luna E, Maggi C, Owsiak M and Scannell R 2021 Pedestal structure, stability and scalings in JET-ILW: the EUROfusion JET-ILW pedestal database *Nucl. Fusion* **61** 016001

[18]   Li K, Li G Q, Zang Q, Zhang T, Liu H Q, Xiang H M, Li Y Y, Wu M F, Wu M Q, Jian X, Li G S, Li H and Snyder P B 2020 Study of H-mode pedestal predictive model on EAST tokamak *Plasma Phys. Control. Fusion* **62** 115007

[19]   Li K, Lao L L, Li G Q, McClenaghan J, Jian X, Osborne T, Evans T E, Smith S P, Meneghini O M and Snyder P B 2021 Study of H-mode pedestal model for helium plasmas in DIII-D *Nucl. Fusion* **61** 096002

[20]   Miller R L, Lin-Liu Y R, Turnbull A D, Chan V S, Pearlstein L D, Sauter O and Villard L 1997 Stable equilibria for bootstrap-current-driven low aspect ratio tokamaks *Phys. Plasmas* **4** 1062–8

[21]   Oyama N, Sakamoto Y, Isayama A, Takechi M, Gohil P, Lao L ., Snyder P ., Fujita T, Ide S, Kamada Y, Miura Y, Oikawa T, Suzuki T, Takenaga H, Toi K and Team  the J-60 2005 Energy loss for grassy ELMs and effects of plasma rotation on the ELM characteristics in JT-60U *Nucl. Fusion* **45** 871–81

[22]   Kirk A, Muller H W, Wolfrum E, Meyer H, Herrmann A, Lunt T, Rohde V and Tamain P 2011 Comparison of small edge-localized modes on MAST and ASDEX Upgrade *Plasma Phys. Control. Fusion* **53** 095008

[23]   Wang Y F, Wang H Q, Xu G S, Jia G Z, Turco F, Petty C C, Chen J L, Yan N, Yang Q Q, Wang L, Chen R, Hu G H, Osborne T H, Snyder P B, Garofalo A M, Gong X Z, Qian J P, Li G Q, Guo H Y and Wan B N 2021 Grassy ELM regime at low pedestal collisionality in high-power tokamak plasma *Nucl. Fusion* **61** 016032